\documentclass[12pt,epsf]{article}

\usepackage{verbatim}

\usepackage{dsfont}
\usepackage{sidecap}
\sidecaptionvpos{figure}{t}
\usepackage{subfigure}

\usepackage[english]{babel}

\usepackage{amssymb,amsmath}
\usepackage{graphicx, xcolor, varwidth}
\usepackage{setspace}
\usepackage[permil]{overpic}
\usepackage{cite}

\colorlet{darkblue}{blue!70!black}

\usepackage[colorlinks=true,urlcolor=darkblue,linktocpage=true,linkcolor=darkblue,citecolor=darkblue]{hyperref}

\numberwithin{equation}{section}

\newcommand{\be}{\begin{equation}}
\newcommand{\ee}{\end{equation}}
\newcommand{\bea}{\begin{eqnarray}}
\newcommand{\eea}{\end{eqnarray}}
\newcommand{\bear}{\begin{eqnarray}}
\newcommand{\eear}{\end{eqnarray}}  
\newcommand{\beas}{\begin{eqnarray*}}
\newcommand{\p}{\partial}
\newcommand{\eeas}{\end{eqnarray*}}
\newcommand{\ba}{\begin{array}}
\newcommand{\ea}{\end{array}}

\newcommand{\tr}{\operatorname{tr}}
\newcommand{\pd}[2][1]{\ifnum#1=1 \frac{\partial}{\partial {#2}} \else
  \frac{\partial^#1}{\partial {#2}^{#1}}\fi}
\newcommand{\dpd}[2][1]{\ifnum#1=1 \dfrac{\partial}{\partial {#2}} \else
  \frac{\partial^#1}{\partial {#2}^{#1}}\fi}
\newcommand{\td}[2][1]{\ifnum#1=1 \frac{d}{d{#2}} \else
  \frac{d^#1}{d{#2}^{#1}}\fi}

\newcommand{\nbox}{{\,\lower0.9pt\vbox{\hrule \hbox{\vrule height 0.2 cm \hskip 0.19 cm \vrule height 0.2 cm}\hrule}\,}}
\newcommand{\Tr}{\ {\rm Tr}\ }

\newcommand{\ie}{{\it i.e.,}\ }

\newcommand{\half}{\tfrac{1}{2}}



\begin{document}

\begin{spacing}{1.3}
\begin{titlepage}

\begin{center}
{\Large \bf Speed Limits for Entanglement\\ \vspace{.3cm}  }

\vspace*{6mm}

Nima Afkhami-Jeddi and Thomas Hartman
\vspace*{6mm}

\textit{Department of Physics, Cornell University, Ithaca, New York\\}

\vspace{6mm}

{\tt na382@cornell.edu, hartman@cornell.edu}

\vspace*{6mm}
\end{center}
\begin{abstract}

We show that in any relativistic  system, entanglement entropy obeys a speed limit set by the entanglement in thermal equilibrium.  
The bound is derived from inequalities on relative entropy with respect to a thermal reference state. Thus the thermal state constrains far-from-equilibrium entanglement dynamics whether or not the system actually equilibrates, in a manner reminiscent of fluctuation theorems in classical statistical mechanics. A similar shape-dependent bound constrains the full nonlinear time evolution, supporting a simple physical picture for entanglement propagation that has previously been motivated by holographic calculations in conformal field theory. We discuss general quantum field theories in any spacetime dimension, but also derive some results of independent interest for thermal relative entropy in 1+1d CFT.

\end{abstract}

\end{titlepage}
\end{spacing}

\vskip 1cm

\setcounter{tocdepth}{2}
\tableofcontents

\begin{spacing}{1.3}
 
\section{Introduction}

Closed quantum systems remain forever in a pure state, but finite subsystems appear to thermalize. This is due to the creation and spreading of entanglement.   In a weakly interacting theory, entanglement is created by interactions and carried by quasiparticles, so there is a simple physical picture for how entanglement entropy of a subsystem increases. In strongly interacting systems, this intuition breaks down, but entanglement generation is nonetheless subject to simple laws.

These laws, studied mostly in lattice systems (for example \cite{sie,mbl}), formalize the idea that even at strong coupling, entanglement is created by interactions and carried by matter.  In this paper we derive a new class of bounds on entanglement generation in relativistic quantum field theories, motivated by similar intuition. (See \cite{Avery:2014dba} for a more direct analogue of the lattice results.)

Almost all known results for time-dependent entanglement entropy in strongly coupled QFT are for quenches in 1+1d conformal field theory, or in higher dimensional CFTs with holographic duals.  In all cases, starting in a translation-invariant initial state, there is a universal regime of linear growth, which can be used to define an effective velocity of entanglement:
\be
v_E \equiv \frac{1}{\mbox{area}(A)}\times\frac{dS_A/dt}{s_{th}(\beta)}   \ ,
\ee
where $S_A$ is the entanglement entropy of region $A$, and $s_{th}(\beta)$ is the entropy density in thermal equilibrium, at inverse temperature $\beta$ defined to have the same energy density as the pure state. In 2d CFT, $v_E=1$, so entanglement propagates as if it were carried by free particles moving at the speed of light \cite{Calabrese:2005in}.  However, in a $d$-dimensional CFT with a gravity dual \cite{Hartman:2013qma}
\be
v_E^{holographic} =\frac{\sqrt{d} (d-2)^{\frac{1}{2}-\frac{1}{d}}}{(2 (d-1))^{1-\frac{1}{d}}} \ .
\ee
This number lies between $\half$ and 1.
The linear behavior was first observed numerically \cite{AbajoArrastia:2010yt,Albash:2010mv};  later $v_E$ was derived analytically from the geometry of black hole interiors \cite{Hartman:2013qma}, and  studied in more generality in \cite{Liu:2013iza} where it was dubbed the `tsunami' velocity.  Despite the name, $v_E$ is not actually a velocity, and it is not obvious that it must obey $v_E<1$.\footnote{What \textit{is} obvious is only that the total time to equilibrate must be at least the light-crossing time of region $A$, so the time-averaged entanglement velocity over this process is bounded by 1 times a geometric factor. However, this average is shape-dependent and generally lower than $v_E^{max}$.}

We will derive the instantaneous speed limit
\be\label{speedlimit}
|v_E| \leq 1 \ ,
\ee 
by a simple argument involving the monotonicity of relative entropy with respect to a thermal reference state.  We assume that the state is translation invariant, but do not assume any particular type of quench state; the absolute value means the bound applies to states evolving toward or away from equilibrium, over distance scales $L \gg \beta$ where $\beta$ is the effective inverse local temperature. Boundary states in conformal field theory, often used to model a critical quench \cite{Calabrese:2005in}, provide one particular example.

On the one hand, it is surprising that an equilibrium quantity $s_{th}(\beta)$ appears at all in a bound on the short-timescale dynamics of states arbitrarily far from equilibrium. On the other hand, it supports the intuition that the thermal state is `as mixed as possible' given some fixed energy density and accessible degrees of freedom, so that even on short time and distance scales, it provides an upper bound on entanglement production. The equilibrium state plays a similar role in the derivation of fluctuation theorems for classical systems forced far from equilibrium \cite{ftreview}.  

The thermal relative entropy also bounds the nonlinear time evolution of entanglement entropy, in a way that depends on the shape of the entangling surface. This qualitatively supports the tsunami picture of \cite{Liu:2013iza} for nontrivial shapes, not necessarily for the value of entanglement entropy but at least as an upper bound. All of our results are based on the observation that thermal relative entropy in a translation-invariant state is simply
\be
S_{rel} \approx s_{th}(\beta) V_A - S_A \ ,
\ee
where $V_A$ is the volume of region $A$, and this equation applies only to extensive, finite contributions as described below. 
Inequalities for relative entropy then relate the entanglement entropies and thermal entropies of different spacetime regions.

Most of our discussion assumes region $A$ has size $L\gg \beta$. In 2d CFT, we compute the modular Hamiltonian exactly, and use the result to extend the arguments to arbitrary $L/\beta$. The exact formula for the thermal modular Hamiltonian in 2d CFT is also likely to be useful in other contexts.


After this work was completed, a different derivation of the instantaneous speed limit $v_E <1$ for half-spaces was found in \cite{Casini:2015zua}. They also explore the role of interactions in increasing $v_E$ over the value for free-streaming particles.  Our derivation allows for an immediate generalization to stronger, shape-dependent bounds, and assigns a simple physical interpretation to the `entanglement tsunami': the area of a region not-yet-reached by the tsunami is equal to the thermal relative entropy. It may also be more easily generalized to inhomogeneous states as discussed further below.

\section{Instantaneous speed limit}

\subsection{Thermal relative entropy}\label{ss:srel}
Consider a relativistic QFT in $d$ spacetime dimensions, in a (possibly mixed) state with density matrix $\rho$. We assume only that the state is translation invariant, with constant energy density $\epsilon$. The entanglement entropy of a spatial subregion $A$ is the von Neumann entropy of the reduced density matrix $\rho_A = \tr_{\bar{A}} \rho$ (where $\bar{A}$ denotes the complement). Our basic tool, inspired by the methods of \cite{grover}, is the relative entropy with respect to a thermal state:
\be
S_{rel}(A) \equiv S(\rho_A | \rho_A^\beta)  \equiv  \Tr \rho_A \log \rho_A - \Tr \rho_A \log \rho_A^\beta \ ,
\ee
where $\rho^\beta = \frac{e^{-\beta H}}{Z(\beta)}$ is the density matrix of a thermal state, with the inverse temperature $\beta=\beta(\epsilon)$ chosen to have energy density $\epsilon$. This can be rewritten as
\begin{align}
S_{rel}(A)=S(\rho_A^\beta)-S(\rho_A)+\langle K_A\rangle -\langle K_A\rangle_\beta \ ,
\label{relent}
\end{align}
where $K_A $ is the modular Hamiltonian of the thermal state, defined by $\rho_A^\beta = \frac{e^{-K_A}}{\Tr e^{-K_A}}$. Here $\langle \cdot \rangle = \Tr \rho \cdot$, $\langle \cdot \rangle_\beta = \Tr e^{-\beta H}\cdot$, and $S(\sigma) = - \Tr \sigma \log \sigma$. Individual terms in \eqref{relent} are UV divergent in quantum field theory, but the relative entropy is finite \cite{Casini:2008cr}.  

A large subregion of a thermal state is itself approximately thermal.  That is, for volume $V_A \gg \beta^{d-1}$,
\be\label{betat}
\rho_A^\beta \approx \frac{e^{-\beta H^{(A)}}}{\Tr e^{-\beta H^{(A)}}}
\ee 
where
\be
H^{(A)} = \frac{1}{2\pi}\int_A T_{00}
\ee
is the ordinary Hamiltonian projected onto region $A$. This follows from locality of the Euclidean path integral on $R^{d-1}\times S^1_{\beta}$ that prepares the thermal state.\footnote{See also a stronger version of this statement explored in \cite{Garrison:2015lva}.} The projection involves an arbitrary truncation at the edges, so \eqref{betat} makes sense only for computing extensive quantities. We will prove the relation \eqref{betat} below for the special case of 2d CFT, where even the edge effects can be taken into account, but it holds in general. Therefore the modular Hamiltonian is simply $K_A = \beta H^{(A)}$, and the extensive part of the thermal entropy is
\be
\hat{S}(\rho_A^\beta) \approx  s_{th}(\beta) V_A \ ,
\ee
where $s_{th}$ is the thermal entropy density. The hat indicates that we keep only the volume term in the thermal entanglement entropy (which automatically discards all UV divergences).

Returning to the relative entropy, the energy terms in \eqref{relent} cancel by design, leaving
\be\label{nicerel}
S_{rel}(A) = s_{th}(\beta) V_A -\hat{S}(\rho_A)  \ ,
\ee
for a region $A$ with length scale $L \gg \beta$.
The same relation was used in \cite{grover} to show that $\hat{S}_A$ is bounded by the thermal value. The hat on $\hat{S}(\rho_A)$ again means that we keep only the finite, extensive part. (In a pure state, this includes volume terms $\sim L^{d-1}$ as well as other terms of comparable size, such as $\xi L^{d-2}$ where $\xi\gg\beta$ is a length scale associated to the state. For example, a quench at $t=0$ leads to a state with the length scale $\xi\sim t$, and we include terms growing as $tL^{d-2}$ in $\hat{S}$ for times $t \gg \beta$.)

In a non-interacting theory of quasiparticles with only pairwise entanglement, there is a simple intuitive picture for \eqref{nicerel}: $\rho$ is a sea of Bell pairs, and the relative entropy counts the number of entangled pairs such that both particles reside in region $A$.

\subsection{Monotonicity}

\begin{figure}\centering
\includegraphics{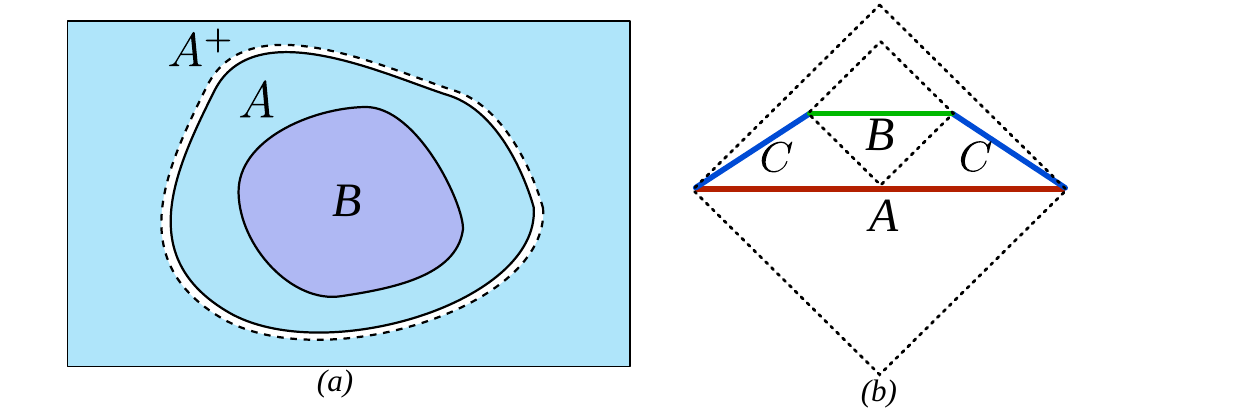}
\caption{\small \textit{(a)} Subregion at a fixed time, $B \subset A$. \textit{(b)} Regions at different times with nested causal diamonds, $D(B) \subset D(A)$.\label{f:subregionsAB}}
\end{figure}

Relative entropy is non-negative, and it is non-increasing under partial trace.  This means that for a subregion $B \subset A$, we have
\be\label{relmono}
S_{rel}(B) \leq S_{rel}(A) \ .
\ee
In translation-invariant states with finite energy density, using \eqref{nicerel} this becomes
\be\label{genbound}
\hat{S}_A - \hat{S}_B < s_{th}(\beta)(V_A - V_B) \ .
\ee
Applied to a fixed time slice as in figure \ref{f:subregionsAB}$a$, this gives an upper bound on $\hat{S}_A$. Strong subadditivity also gives a lower bound, as follows. Define the region $A^+$ to be the complement of a region slightly larger than $A$, as in figure \ref{f:subregionsAB}$a$. (This provides a convenient regulator for the entanglement entropy via the mutual information, $I(A|A^+) \approx 2 \hat{S}_A$ \cite{Casini:2015woa}. In our case we can separate the boundaries of $A$ and $A^+$ by $\sim \beta$.) Choosing $X=A, Y=BA^+$, SSA in the form $S_X + S_Y \geq S_{X\cup Y} +  S_{X \cap Y}$ leads to a triangle inequality for the finite parts.  Together with the relative entropy bound, we have
\be\label{sandwich}
|\hat{S}_A - \hat{S}_B| \leq s_{th}(\beta)\Delta V
\ee
where $\Delta V = V_{A\backslash B}$. In words, this is intuitive: by adjoining matter to system $B$, we can entangle or disentangle the system at most by the amount of matter added. 

In a relativistic theory, relative entropy depends only on the causal diamond $D(A)$ associated to a region $A$, not on a choice of time slicing. Accordingly,  monotonicity \eqref{relmono} holds more generally for nested causal diamonds,
\be
D(B) \subset D(A) \ .
\ee
The reason is simple: referring to figure \ref{f:subregionsAB}$b$, the usual monotonicity inequality implies $S_{rel}(B) \leq S_{rel}(BC)$, and Lorentz invariance implies $S_{rel}(BC) = S_{rel}(A)$.\footnote{Roughly speaking, this last equality follows from the fact that there is a unitary evolution connecting any two slicings of the same causal diamond, $\rho_{BC} = U^{\dagger}\rho_A U$. This argument ignores UV divergences, which can be addressed in the language of operator algebras without changing the conclusion; see \cite{haag} and the remarks in \cite{Casini:2008wt,Blanco:2013lea}.}

\begin{figure}\centering
\includegraphics{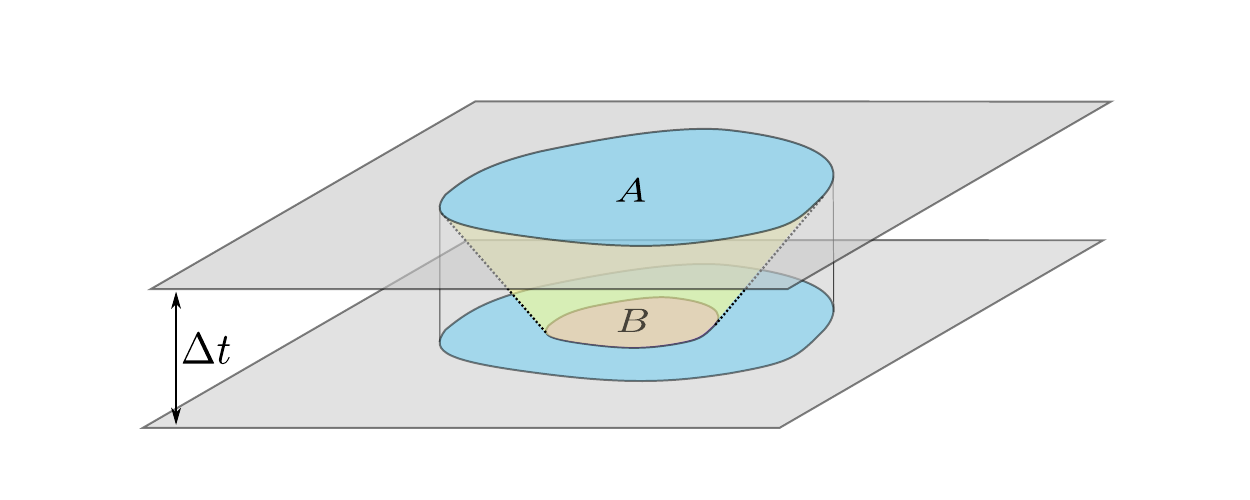}
\caption{\small Setup for the derivation of speed limits. The causal diamonds are nested, $D(B) \subset D(A)$, and the boundary of $B$ is null separated from the boundary of $A$.\label{f:nicediamond}}
\end{figure}

The starting point for our dynamical constraints is monotonicity \eqref{genbound} applied to the configuration in figure \ref{f:nicediamond}. Region $A$ is at time $t$, and region $B$ is defined at time $t - \Delta t$ by sending null rays back in time from the boundary of $A$. 

 \subsection{Speed limit for a strip}\label{ss:vhalf}
\begin{figure}
\includegraphics{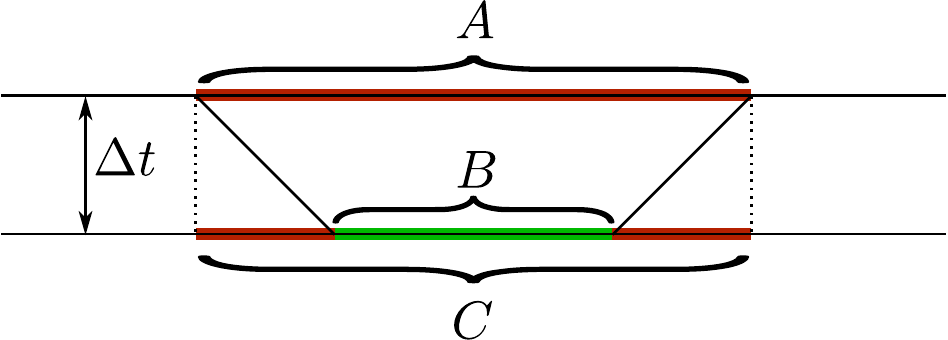}\centering
\caption{Regions for the derivation of the speed limit for a strip.\label{f:strip}}
\end{figure}

Consider a strip extending a distance $L$ in the $x$ direction and infinitely in the other spatial directions. Denote the entanglement entropy (divided by the transverse volume) by $\hat{S}_{strip}(L,t)$.  It was shown in \cite{grover} that in a translation-invariant state, this entropy increases monotonically with size:
\be\label{monol}
\p_L \hat{S}_{strip}(L,t) \geq 0  \ .
\ee
 The argument of \cite{grover} is reviewed in appendix A. To derive a speed limit, we now apply \eqref{genbound} as shown in figure \ref{f:strip} with infinitesimal $\Delta t$. Using monotonicity under partial trace and \eqref{monol},
\be\label{abc}
\hat{S}_A \leq \hat{S}_B + s_{th}(\beta)(V_A-V_B) \leq \hat{S}_{C} + s_{th}(\beta)(V_A - V_B) \ ,
\ee
which implies the upper speed limit $\frac{\p}{\p t} \hat{S}_{strip}(L,t) \leq 2 s_{th}(\beta)$.
Time-reversing the figure gives an identical bound for the rate of decrease, so we find the speed limit
\be
\left|\frac{\p}{\p t} \hat{S}_{strip}(L,t) \right| \leq 2 s_{th}(\beta)  \ .
\ee
The factor of two is the area divided by transverse volume, so this is the bound $|v_E|\leq 1$ for strips.

\subsection{General shapes}
The only fact about the strip geometry important for this derivation was that entropy increases with size at fixed time, $\p_L \hat{S}_A \geq 0$. In the appendix we show that 
in translation-invariant pure states,
this holds for arbitrary convex shapes, up to small corrections that do not contribute to the finite, extensive term $\hat{S}_A$. Therefore, we have the speed limit
\be\label{gspeed}
\left| \frac{\p}{\p t}\hat{S}_A(t) \right| \leq s_{th}(\beta) \times \mbox{area}(A) \ ,
\ee
for convex regions.
As described in the introduction, this is the speed limit $|v_E| \leq 1$ on the tsunami velocity in the sense of \cite{Hartman:2013qma,Liu:2013iza}. This bound is curiously large, as explored in \cite{Casini:2015zua}: it is easy to construct quasiparticle states in which $v = 1$ in some particular direction, but not in all directions. Isotropic free streaming leads to a significantly lower velocity. Conformal field theories with holographic duals produce velocities with $v_{\rm free\  streaming} < v_E < 1$, and it is not known whether $v_E = 1$ can actually be approached or saturated.


\section{Shape dependence}
The instantaneous speed limit depends only on the thermal entanglement entropy and the surface area of the region.  A similar argument leads to more stringent, shape-dependent bounds on the full function $\hat{S}_A(t)$. For concreteness, we choose $A$ to be a ($d-1$)-cube with sides of length $L$ and consider a quench experiment where the system starts in a gapped state, then approaches equilibrium. That is,
\be
\hat{S}_{cube}(L, t=0) = 0 \ ,
\ee
and the saturation time $t_{sat}(L)$ is defined as the earliest time for which
\be
\hat{S}_{cube}(L, t_{sat}) = s_{th}(\beta) V_L \ .
\ee
Now we apply \eqref{genbound} to the configuration in figure \ref{f:nicediamond} (or its time-reverse), optimizing $\Delta t$ to produce the strongest constraint. It turns out that the strongest constraint for an upper bound comes from choosing region $B$ at $t=0$, and the strongest lower bound comes from the time-reversed figure with $B$ at $t=t_{sat}$. Combining these two constraints gives nonlinear bounds on the time dependence throughout the thermalization process.  

\begin{figure}
\begin{center}
\includegraphics[scale=1]{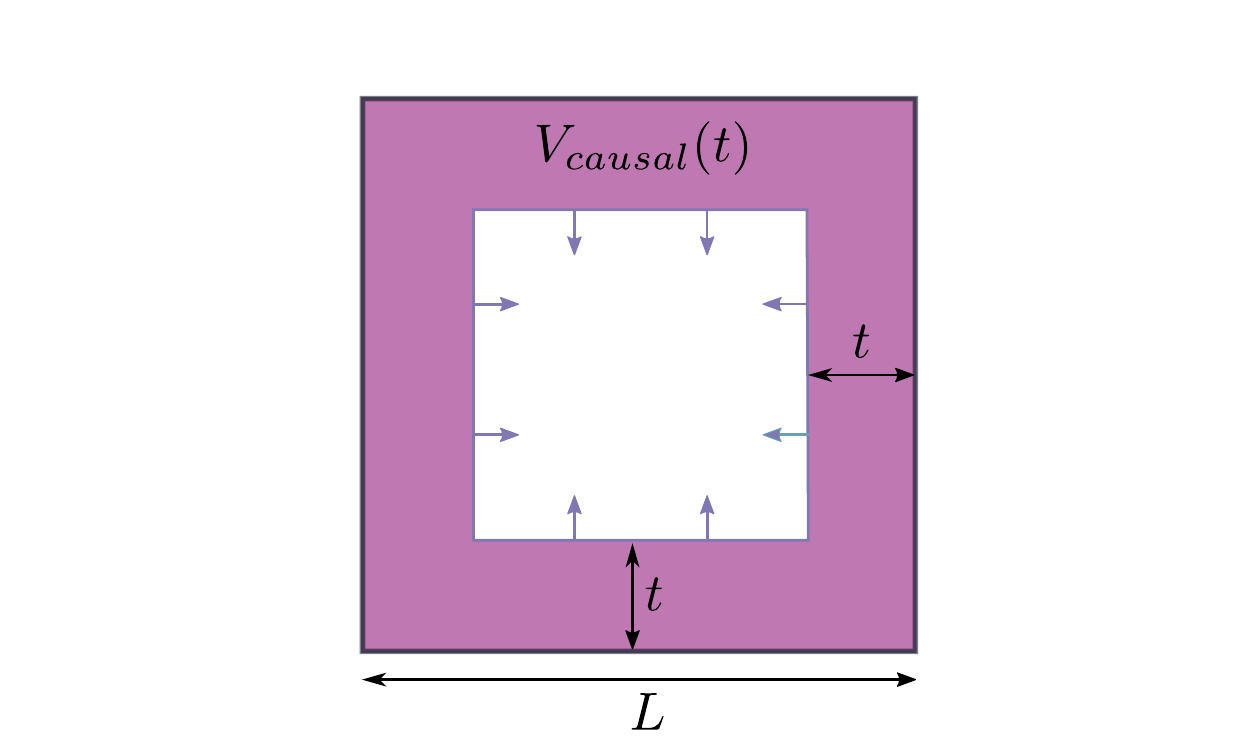}
\end{center}
\caption{The causal volume $V_{causal}(t)$ is defined as the volume of region $A$ at time $t$ causally connected to the region $\bar{A}$ at $t=0$.\label{f:causalv}}
\end{figure}

To state the bound, define the causal volume $V_{causal}(t)$ as the volume of the cube causally connected to the exterior after a time $t$.  This volume, illustrated in figure \ref{f:causalv},  is identical to the concept of the `entanglement tsunami' proposed in \cite{Liu:2013iza} based on a qualitative picture for entanglement spreading. The upper and lower bounds are then
\be\label{cubecausal}
s_{th}V_{causal}^{cube}(t_{sat}-t) <  \hat{S}_{cube}(L, t) < s_{th} V_{causal}^{cube}(t) \ ,
\ee
where
\be
V_{causal}^{cube} (t) = \left\{ \begin{array}{cc}
L^{d-1} - (L-t)^{d-1} & t<L/2\\
L^{d-1} & t>L/2
\end{array}\right. \qquad .
\ee
The same logic can be applied to any shape for which $\p_L S_A \geq 0$.  For example, the entanglement entropy of a disk in $2+1$ dimensions after a quench is bounded by an equation like \eqref{cubecausal} with
\be
V_{causal}^{disk}(t) = \left\{ \begin{array}{cc}
\pi t (2R-t)& t<R\\
\pi R^2 & t>R
\end{array}\right. \qquad .
\ee

\section{Conformal field theory in 1+1 dimensions}\label{s:cft}
Our results are simply illustrated in 2d conformal field theory, a setting in which quenches have been studied extensively both in field theory \cite{Calabrese:2005in} and in holographic models \cite{AbajoArrastia:2010yt,Albash:2010mv,Balasubramanian:2010ce,Hartman:2013qma}.

\begin{figure}
\begin{center}
\includegraphics{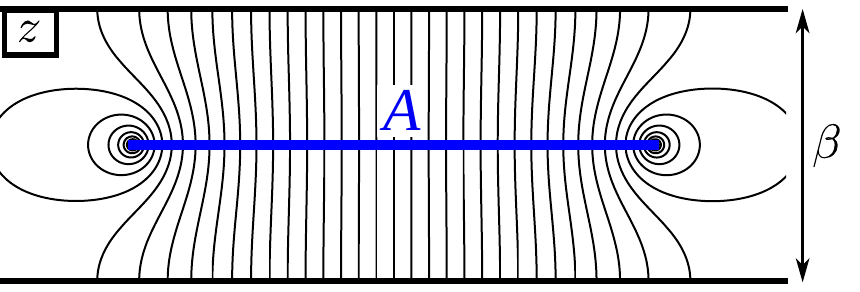}
\end{center}
\caption{Vector field on the Euclidean thermal cylinder which, upon continuation to Lorentzian signature, generates the modular evolution of an interval in a thermal state.\label{f:zeta}}
\end{figure}

First, we revisit the discussion of the modular Hamiltonian that led to the thermal relative entropy \eqref{nicerel}. There we used locality to relate the modular Hamiltonian to the projected Hamiltonian for large $L$. But in 2d CFT, the modular Hamiltonian $K_A$ can be computed exactly, for arbitrary $L$. It is the conserved charge associated to a conformal boost vector $\zeta$ shown in figure \ref{f:zeta} on the Euclidean thermal cylinder:
\be\label{kch}
K_A = \int_A dV n^\mu T_{\mu\nu} \zeta^\nu \ ,
\ee
where $T$ is the stress tensor and $n = \hat{t}$ is the timelike unit normal.  To find $\zeta$ explicitly, choose a complex coordinate $z$ on the thermal cylinder with periodic identification $z \sim z + i \beta$. Region $A$ is the segment $\mbox{Re\ }z \in [0, L]$ at Im $ z = 0 $.
The conformal mapping $z=\frac{2\pi}{\beta}\log\left(\frac{1+e^{\frac{2\pi L}{\beta}} w}{1+w}\right)$ takes this region to the half-line $w \in [0, \infty]$ on the $w$-plane.  Writing $w = r e^{i \phi}$, the rotational vector field $\zeta = \hat{\phi}$ generates the modular evolution of region $A$; this is the familiar statement that the modular Hamiltonian of Rindler space is the boost charge (see for example \cite{Casini:2011kv}). This flow vector can then be translated back into cylinder coordinates $z = x + i \tau$. Plugging into \eqref{kch} gives the exact modular Hamiltonian of the region $x \in [0,L]$ in a thermal state:
\be
K_A = \frac{\beta}{\pi} \int_0^L dx\frac{\sinh\left(\pi x \over \beta\right) \sinh \left( \pi(L-x)\over \beta\right)}{\sinh\left( \pi L \over \beta\right)} T_{00}(x) \ .
\ee
For distance scales $x,L,L-x \gg \beta$, the prefactor is a step function. Therefore, up to boundary terms, the modular Hamiltonian is indeed the ordinary Hamiltonian projected onto region $A$:
\be
K_A \approx \frac{\beta}{2\pi}\int_0^L dx \, T_{00}(x) \ .
\ee
This confirms the discussion of section \ref{ss:srel} in this context. 

On the other hand, since both the thermal entanglement entropy and modular Hamiltonian are known exactly, we do not need to resort to any approximations to derive a speed limit for 2d CFT.  Since we chose the reference state such that $\langle T_{00} \rangle_\beta = \langle T_{00}\rangle$, the energy terms in thermal relative entropy \eqref{relent} cancel exactly -- \ie keeping all terms, including non-extensive contributions.  The exact thermal entanglement entropy is \cite{Korepin:2004zz,Calabrese:2004eu}
\begin{align}
S(\rho_A^\beta) =\frac{c}{3}\log\left[\frac{\beta}{\pi \epsilon_{UV}}\sinh\left(\frac{ \pi L }{\beta}\right)\right],
\end{align}
where $\epsilon_{UV}$ is the UV cutoff. Repeating the logic of section \ref{ss:vhalf}, now without any approximations, we find
\be
S(L,t+\Delta t) - S(L, t) \leq S_{thermal}(L) - S_{thermal}(L-2\Delta t) \ ,
\ee
where $S(L,t)$ is the entanglement entropy of the region $A = \{ x \in[0,L]\}$ at time $t$.  Sending $\Delta t \to 0$ gives the exact result
\be
\left| \p_t S(L,t) \right| \leq \frac{2\pi c}{3\beta} \coth \left( \pi L \over \beta\right)\ .
\ee
This agrees with the results of section \ref{ss:vhalf} for $L\gg \beta$, but is weaker for $L \sim \beta$. Presumably, the weaker bound for $L \lesssim \beta$ is because the thermal ensemble does not provide a good reference state in this regime. 

\section{Discussion}
Our main message is that thermal relative entropy provides a simple tool to turn intuition about entanglement --- for example, that it is carried by matter, and increases or decreases as matter spreads --- into theorems for strongly interacting systems. We have given several examples in relativistic QFT, and expect that there are other similar applications. It would be interesting to apply similar tools to lattice systems, and perhaps draw a connection to the small incremental entangling theorem \cite{sie}.  

Finally, it may be useful to generalize these methods to study inhomogeneous states.  The obvious choice for a reference state is the classical hydrodynamic state with the same value of $\langle T_{\mu\nu}\rangle$. Just as the thermal state bounds the entanglement dynamics of arbitrary, non-equilibrium translation-invariant states, we expect classical hydrodynamics to constrain the dynamics of highly quantum states with nontrivial energy transport.

\bigskip

\bigskip

\noindent  \textbf{Acknowledgments} 

\noindent We are grateful to Curtis Asplund, Alice Bernamonti, Federico Galli, James Garrison, Tarun Grover, Hong Liu, and Mark Mezei for useful discussions,  Venkatesa Chandrasekaran and Amir Tajdini for discussions and collaboration on related work, and Tarun Grover for comments on a draft.  This work is supported by DOE Early Career Award DE-SC0014123. TH also thanks the KITP for support during the program ``Entanglement in Strongly-Correlated Quantum Matter" under NSF PHY11-25915. NA is also supported by is supported by Natural Sciences and Engineering Research Council of Canada.

\appendix

\section{Entropy increases with region size}

\begin{figure}
\begin{center}
\includegraphics[scale=1]{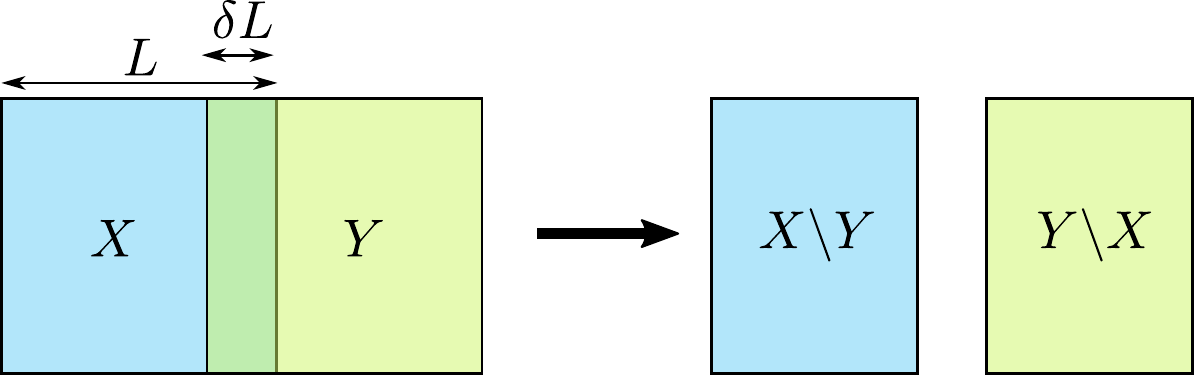}
\end{center}
\caption{Strong subadditivity applied to two rectangles implies that the entanglement entropy of the rectangle $X$ is larger than the entanglement entropy of the smaller rectangle $X\backslash Y$\label{f:rectangles}}
\end{figure}

\begin{figure}
\begin{center}
\includegraphics[scale=1]{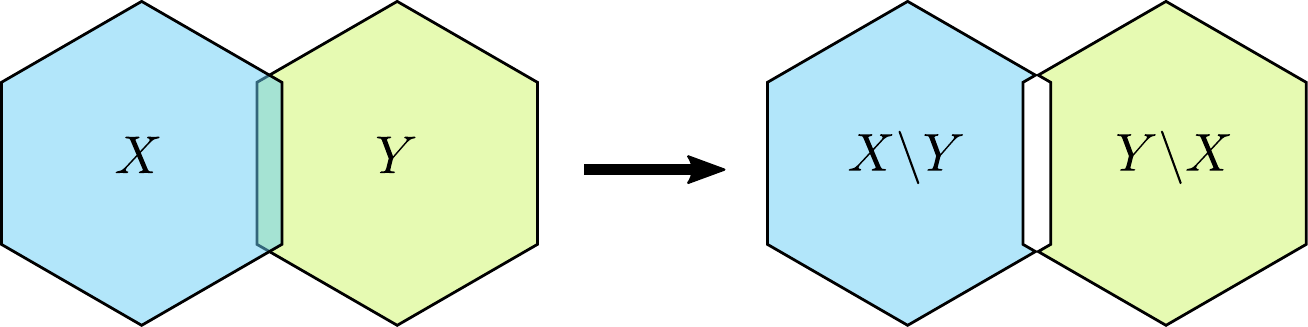}
\end{center}
\caption{Strong subadditivity applied to two regular polygons results in a smaller polygon plus extra corner contributions that do not scale with $L$ implying the  entanglement decreases with decreasing size. \label{f:polygons}}
\end{figure}

\begin{figure}
\begin{center}
\includegraphics[scale=1]{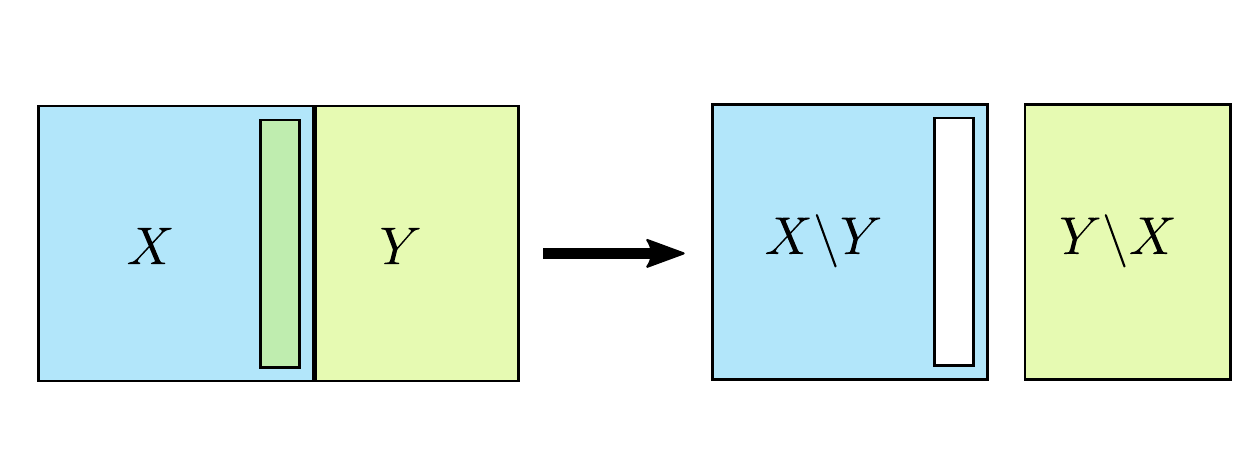}
\end{center}
\caption{UV finite version of SSA (see \eqref{uvssa}) \label{f:newrectangles}}
\end{figure}

In this appendix we show that for convex regions, the regulated entanglement entropy $\hat{S}_A$ in a translation-invariant state increases monotonically with the size of the region. In this scaling, the shape is held fixed in the sense of figure \ref{f:causalv}. The argument generalizes that for strips in \cite{grover}, using strong subadditivity in the form \cite{Lieb:1973cp}
\be\label{ssa}
S_X + S_Y \geq S_{X\backslash Y} + S_{Y \backslash  X} \ .
\ee
For the strip, choose $X$ to be $x \in [0,L]$, and $Y$ to be $x \in [L-\delta L, 2L - \delta L]$. Then SSA implies $S_{strip}(L) > S_{strip}(L - \delta L)$. This can be generalized to higher-dimensional shapes for which the transformation $G(L+\delta L) \to G(L)$ can be effected by intersecting two copies of the original shape. Ignoring UV divergences for now, the details for a two-dimensional rectangle are explained in figure \ref{f:rectangles}. The same method also applies to convex polygons, up to extra corner contributions illustrated in figure \ref{f:polygons}. The contribution of the corners does not scale with $L$, so it does not contribute to $\hat{S}_A$. Thus for convex polygons we conclude that the entanglement entropy is monotically increasing with $L$. 
However, we have ignored UV divergences, which seem to render this statement trivial: the area divergence increases with $L$, so we cannot conclude anything about the finite terms. To escape this problem we can at every stage use a modified version of \eqref{ssa} that is UV-finite.  Motivated by the mutual information regulator introduced in \cite{Casini:2015woa}, we replace in \eqref{ssa}
\be
\label{uvssa}
Y \to (Y \backslash X) \cup (X \cap Y)^- \ ,
\ee
where the superscript $A^-$ denotes a region slightly inside of $A$, with boundary separated by $\sim \beta$. This ensures that both sides of \eqref{ssa} have the same UV divergences coming from area and corner contributions, as illustrated in figure \ref{f:newrectangles}. This eliminates all UV divergences from the inequalities in the argument above. It also introduces new errors into each step, but these are proportional to powers of $\beta$ so do not contribute to $\hat{S}$. Therefore we have finally
\be
\hat{S}_A(L) \geq \hat{S}_A(L-\delta L)
\ee
for convex polygons.

Next consider an arbitrary smooth convex shape, for example the $(d-1)$-sphere. The idea is to approximate this by a polygon with a large number of sides, and check that the errors in this procedure are smaller than the extensive contributions we are keeping.  Define the distance $D$ to be the maximal distance between our polygonal approximation and the actual entangling surface. The number of faces necessary to achieve a maximal distance $D$ scales as $n \sim (L/D)^{(d-2)/2}$, and the size of each face as $h \sim \sqrt{DL}$. According to \eqref{sandwich}, the error in $\hat{S}_A$ introduced by approximating region $A$ as a polygon is $\mathcal{E}_{poly} \sim \Delta V \sim DL^{d-2}$. We now take the polygon and scale $L \to L - \ell$. By the argument above, the change in entropy of the polygon is
\be\label{polyscale}
\hat{S}_{poly}(L) -\hat{ S}_{poly}(L-\ell)\geq  \mathcal{E}_{scaling}
\ee
where the left-hand side has an extensive contribution $\sim \ell L^{d-2}$, and the error from inexact overlap on the right-hand side $\mathcal{E}_{scaling} \sim n \ell^{d-1}$. Combining \eqref{polyscale} with \eqref{sandwich} and the error estimate for $\mathcal{E}_{poly}$ we have
\be
\hat{S}_A(L) - \hat{S}_A(L-\ell) \geq \mathcal{E}_{scaling} + \mathcal{E}_{poly} \sim (L/D)^{(d-2)/2} \ell^{d-1} + D L^{d-2} \ .
\ee
The leading finite terms that we are attempting to bound on the left-hand side scale as $\ell L^{d-2}$. Therefore, in order to drop the error terms on the right-hand size, we need to check that
\be
(L/D)^{(d-2)/2} \ell^{d-1} \ll \ell L^{d-2} \quad , \quad D L^{d-2} \ll \ell L^{d-2} \ .
\ee
That is, $D \ll \ell \ll \sqrt{\ell D}$. It is always possible to choose $\ell,D$ so this holds, for example by setting $D = \epsilon^{3/2} L$ and $\ell = \epsilon L$ with $ \frac{\beta}{L} \ll \epsilon \ll 1$. Therefore,
\be
\hat{S}_A(L) - \hat{S}_A(L-\ell) \geq 0  \ , 
\ee
as claimed.

\end{spacing}

\end{document}